\documentclass[10pt,twocolumn]{article}

\usepackage[a4paper,margin=1in]{geometry}  % Page size and margins
\usepackage{graphicx}  % Figures
\usepackage{xurl}  % Allow line breaks in URLs
\usepackage{hyperref}  % Clickable links
\usepackage[table,xcdraw]{xcolor}  % Colors for tables
\usepackage{colortbl}  % Table color enhancements
\usepackage{booktabs}  % Improved table formatting
\usepackage{caption}  % Better figure captions
\usepackage{amsmath,amssymb}  % Math packages
\usepackage{titlesec}  % Section formatting
\usepackage{tabularx}  % Advanced tables
\usepackage{multirow}  % Multi-row table cells
\usepackage{tcolorbox}  % Colored boxes
\usepackage{listings}  % Code listings
\usepackage{float}  % Better float placement
\usepackage{placeins}  % Prevents figures from floating past sections
\usepackage[numbers]{natbib}  % Proper citation style
\usepackage{fancyhdr}  % Custom headers and footers
\usepackage{authblk}  % Proper author and affiliation formatting
\title{\textbf{Large Language Models for Code Generation: A Comprehensive Survey of Challenges, Techniques, Evaluation, and Applications}}

\author{Nam Huynh}
\author{Beiyu Lin}

\affil{School of Computer Science, University of Oklahoma, Norman, OK, USA}

\date{}  % Removes default date

\begin{document}

\maketitle  % Generates title

\begin{abstract}
Large Language Models (LLMs) have demonstrated their remarkable capabilities in numerous fields. This survey focuses on how LLMs empower users, regardless of their technical background, to use human languages to automatically generate executable code. We begin with understanding LLMs' limitations and challenges in automated code generation. Subsequently, we review various fine-tuning techniques designed to enhance both the performance and adaptability of LLMs in code generation tasks. We then review the existing metrics and benchmarks for evaluations to assess model performance based on fine-tuning techniques. Finally, we explore the applications of LLMs (e.g. CodeLlama, GitHub Copilot, ToolGen) in code generation tasks to illustrate their roles and functionalities. This survey provides a comprehensive overview of LLMs for code generation, helps researchers in diverse fields better understand the current state-of-the-art technologies, and offers the potential of effectively leveraging LLMs for code generation tasks.
\end{abstract}

\vspace{1em} % Adjust spacing as needed
\noindent \textbf{Index Terms—} Large Language Models (LLMs), Code Generation, Machine Learning, Artificial Intelligence (AI).

\section{Introduction}
In general, data mining requires users with good coding skills and domain knowledge based on extensive training hours \cite{discoverdatascience}. A competent data mining expert needs a foundational education in computer science or related fields along with practical experience in programming languages (PLs), such as Python and R \cite{edx}. Coding with a programming language like Python and its powerful libraries like Pandas can make these efforts easier by automating many preprocessing tasks \cite{freecodecamp}. However, for example, data preprocessing, a critical step in data mining, is still very time consuming. The New York Times states that data preprocessing accounts for 50–80\% of a data scientist's workload \cite{freecodecamp}.  But now, with assistance from LLMs, it would only take several minutes to automatically generate code for data preprocessing. Figure \ref{chatgpt_example} shows an example of using ChatGPT to automatically generate code for a data preprocessing task (replacing missing values). The accessibility, effectiveness, and efficiency of using LLMs in daily tasks show the huge potential of applying LLMs in automatic code generation. We write this survey in the hope of opening the possibilities of Data Mining for everyone.

\begin{figure}[htbp]
    \centering
    \begin{tcolorbox}
      
        \textbf{User Prompt:}
        \begin{tcolorbox}[colback=blue!5!white, colframe=blue!50!black, boxrule=0.25mm, rounded corners, fontupper=\normalsize]
        Help me generate code for the following task: Remove rows with missing Product\_ID, replace missing Price values with the median, forward-fill missing Quantity values, and save the cleaned dataset as cleaned\_sales\_data.csv.

        In addition, record the time you take to generate this code.
        \end{tcolorbox}

        \textbf{ChatGPT 4o Output: Python Code}
        \begin{tcolorbox}[colback=green!5!white, colframe=green!50!black, boxrule=0.25mm, rounded corners, fontupper=\normalsize]
        \begin{lstlisting}[language=Python, basicstyle=\ttfamily\normalsize] % Normalsize font matches other boxes

# Import pandas library        
import pandas as pd

# Load the dataset
data = pd.read_csv('sales_data.csv')

# Remove rows with missing Product_ID
data = data.dropna(subset=['Product_ID'])

# Replace missing Price values with the median
median_price = data['Price'].median()
data['Price'] = data['Price'].fillna(median_price)

# Forward-fill missing Quantity values
data['Quantity'] = data['Quantity'].fillna(method='ffill')

# Save the cleaned dataset
data.to_csv('cleaned_sales_data.csv', index=False)

print("Data preprocessing completed successfully.")
        \end{lstlisting}
        \end{tcolorbox}

        \textbf{Time Taken:}
        \begin{tcolorbox}[colback=yellow!5!white, colframe=yellow!50!black, boxrule=0.25mm, rounded corners, fontupper=\normalsize]
        The code generation took approximately 0.000044 seconds (44 microseconds).
        \end{tcolorbox}

    \end{tcolorbox}
    \caption{Example of using ChatGPT 4o to generate code for removing missing values.}
    \label{chatgpt_example}
\end{figure}

Today, it is feasible to train a Generative AI (GenAI) model based on vast amounts of rich and diverse data from various resources, such as code repositories, technical forums, and web data on coding aspects \cite{fabrichq}. For example, a descendant of GPT-3, OpenAI Codex, has been trained on data from billions of source code, such as code in public GitHub repositories \cite{openaicodex}. The comprehensive training on rich data enables LLMs to better understand the context of code comments and function names, as well as to better interpret variable names \cite{fabrichq}. For example, since LLMs have also been integrated into Integrated Development Environments (IDEs), such as PyCharm and VSCode to allow programmers to develop their code, LLMs within these environments can comprehend this contextual information and provide suggestions for users \cite{vectara}. Figure \ref{copilot_example} shows an example of using GitHub Copilot to handle a data preprocessing task - splitting the dataset \cite{datacampcopilot}.

\begin{figure}[h!]
      
        \textbf{Given the comment:}
        \begin{tcolorbox}[colback=blue!5!white, colframe=blue!50!black, boxrule=0.25mm, rounded corners, fontupper=\normalsize]
        \begin{lstlisting}[basicstyle=\ttfamily\normalsize, showstringspaces=false]
# Split the data into features and target variable
        \end{lstlisting}
        \end{tcolorbox}

        \textbf{GitHub Copilot generates the following code:}
        \begin{tcolorbox}[colback=green!5!white, colframe=green!50!black, boxrule=0.25mm, rounded corners, fontupper=\normalsize]
        \begin{lstlisting}[language=Python, basicstyle=\ttfamily\normalsize, showstringspaces=false]
X = df.drop(['Date', 'Rented Bike Count'], axis=1)
y = df['Rented Bike Count']
        \end{lstlisting}
        \end{tcolorbox}

    \caption{Example of using GitHub Copilot generating code \cite{datacampcopilot}}
    \label{copilot_example}
\end{figure}

This survey outlines key aspects of LLMs in code generation by organizing them into four parts: (III) limitations and challenges, (IV) fine-tuning techniques, (V) evaluations, and (VI) applications. These four parts are supported by 38 references (including research papers and technical articles). Each part is further detailed by specific subtopics. Part III, Limits and Challenges of LLMs in automatic code generation, has four topics: (A) resource constraints, (B) syntactic and semantic errors, (C) biases, and (D) security risks. Subsequently, Part IV, Fine-tuning Techniques for better performance and adaptability of LLMs, covers topics ranging from simple prompt engineering to more complex approaches, such as reinforcement learning and domain-specific dataset fine-tuning. Next, Part V, Evaluations, introduces established metrics and benchmarks that quantify the effectiveness and reliability of these techniques from the second part. Finally, Part VI, LLMs' applications, examines the practical applications of LLMs in automating specific coding tasks. These tasks represent various aspects of code generation, including foundational tasks like code generation and code completion, to advanced tasks such as advanced code generation and code search, and auxiliary tasks including debugging and code translation. By integrating these findings, this paper consolidates existing knowledge and identifies gaps and opportunities to advance LLMs in code generation. The layout of this survey paper is shown in Table \ref{tab2}, which serves as a roadmap for the discussions presented in the paper.

\renewcommand{\arraystretch}{1.5}
\begin{table*}{\centering}
\centering
\caption{Summary of Reviewed Studies}
\small
\begin{tabular}{|c|c|c|}
\hline
\textbf{Topic} & \textbf{Subtopics} & \textbf{Sources} \\
\hline
\multirow{10}{*}{\textbf{III. Limits and Challenges}}
& \begin{tabular}[c]{@{}c@{}}A. Resource Constraints\end{tabular} 
& \begin{tabular}[c]{@{}c@{}}Training Requirements \cite{Chen} \\ 
Memory Efficiency \cite{Nyun} \\ 
Performance of Smaller Models \cite{Hassid}\end{tabular} \\
\cline{2-3}
& \begin{tabular}[c]{@{}c@{}}B. Syntactic and Semantic Errors\end{tabular} 
& \begin{tabular}[c]{@{}c@{}}Common Issues \cite{Wang} \\ 
Errors Analysis \cite{Dou} \\ 
Errors in Translation \cite{Pan} \\ 
ChatGPT Errors \cite{Liu}\end{tabular} \\
\cline{2-3}
& \begin{tabular}[c]{@{}c@{}}C. Biases\end{tabular} 
& \begin{tabular}[c]{@{}c@{}}Multilingual Bias \cite{bias2} \\ 
Social Bias \cite{bias3}\end{tabular} \\
\cline{2-3}
& \begin{tabular}[c]{@{}c@{}}D. Security Risks\end{tabular} 
& \begin{tabular}[c]{@{}c@{}}Security Vulnerabilities \cite{Islam2024} \\ 
Systematic Testing \cite{He2023} \\ 
Prompting Strategies \cite{Black2024} \\ 
Vulnerable Data Sources \cite{Wang2024}\end{tabular} \\
\hline
\multirow{6.8}{*}{\textbf{IV. Fine-tune Techniques}} 
& \begin{tabular}[c]{@{}c@{}}A. Fine-Tuning on \\ Domain-Specific Datasets\end{tabular} 
& \begin{tabular}[c]{@{}c@{}}Instruction-Tuning for Optimization \cite{llamoco} \\ 
Parameter-Efficient Fine-Tuning \cite{4peft} \\ 
Data Pruning for Fine-Tuning \cite{datapruning}\end{tabular} \\
\cline{2-3}
& \begin{tabular}[c]{@{}c@{}}B. Feedback\end{tabular} 
& \begin{tabular}[c]{@{}c@{}}ClarifyGPT Framework \cite{mu} \\ 
RLEF \cite{gehring} \\ 
Crowd-Sourced RLHF \cite{cRLHF}\end{tabular} \\
\cline{2-3}
& \begin{tabular}[c]{@{}c@{}}C. Prompt Engineering\end{tabular} 
& \begin{tabular}[c]{@{}c@{}}Chain-of-Thought Prompting \cite{sun2024} \\ 
AceCoder Technique \cite{li2024} \\ 
Security-Focused Prompting \cite{5prompt}\end{tabular} \\
\hline
\multirow{5}{*}{\textbf{V. Evaluations}} 
& \begin{tabular}[c]{@{}c@{}}A. Metrics\end{tabular} 
& \begin{tabular}[c]{@{}c@{}}CodeBLEU \cite{Ren2020} \\ 
pass@k \cite{humaneval} \\ 
pass-ratio@n \cite{Yeo2024} \\ 
ICE-Score \cite{zhuo}\end{tabular} \\
\cline{2-3}
& \begin{tabular}[c]{@{}c@{}}B. Benchmarks\end{tabular} 
& \begin{tabular}[c]{@{}c@{}}HumanEval \cite{humaneval} \\ 
ClassEval \cite{classeval} \\ 
SWE-bench \cite{swe} \\ 
BigCodeBench \cite{bigcodebench}\end{tabular} \\
\hline
\multirow{7}{*}{\textbf{VI. LLMs' Applications}} 
& \begin{tabular}[c]{@{}c@{}}A. Code Generation \\ and Code Completion\end{tabular} 
& \begin{tabular}[c]{@{}c@{}}GitHub Copilot \cite{copilot} \\ 
Code Llama \cite{codellama} \\ 
ToolGen \cite{toolgen}\end{tabular} \\
\cline{2-3}
& \begin{tabular}[c]{@{}c@{}}B. Advanced Code \\ Generation and Code Search\end{tabular} 
& \begin{tabular}[c]{@{}c@{}}RepoRift \cite{reporift} \\ 
CodeBERT \cite{codebert} \\ 
AlphaCode \cite{alphacode}\end{tabular} \\
\cline{2-3}
& \begin{tabular}[c]{@{}c@{}}C. Code Debugging \\ and Code Translation\end{tabular} 
& \begin{tabular}[c]{@{}c@{}}GPT-4 \cite{gpt4} \\ 
Codex \cite{codex} \\ 
Flourine \cite{flourine}\end{tabular} \\
\hline
\end{tabular}
\end{table*}

\section{Overview: LLMs for Code Generations}
\subsection{Large Language Models (LLMs)}
\begin{figure}[hb!]
  \centering
  \includegraphics[width=7cm, height=6cm]{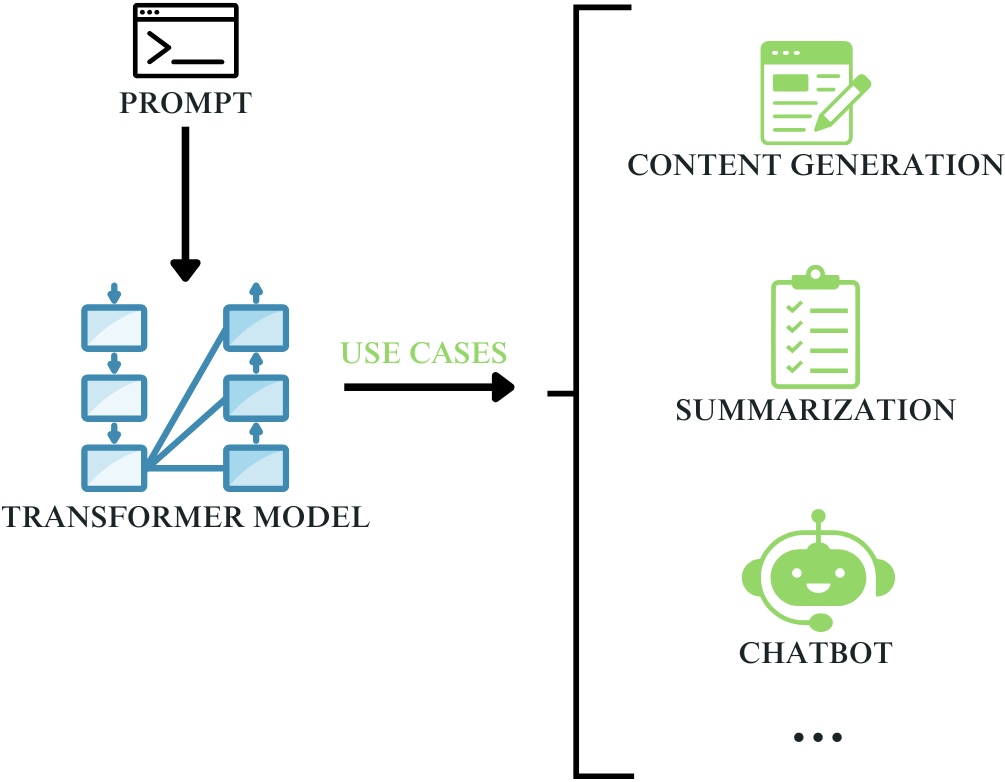}
  \caption{How transformer models work \cite{nvidia1}}
  \label{transformer}
\end{figure}

LLMs have demonstrated significant breakthroughs in many different fields ranging from healthcare to education by using advanced deep learning architectures and training on extensive, diverse datasets \cite{ibmllms}. The deep learning architecture, transformer networks, allows LLMs to understand the meaning of human language \cite{aws}. Transformer networks identify relationships within sequential data, such as words in a sentence \cite{nvidia1}. A transformer consists of multiple layers, such as self-attention layers and feed-forward layers \cite{nvidia1}. These layers can be stacked to create deeper transformers and develop more powerful models. Figure \ref{transformer} is an illustration of how transformer models work. For example, as the transformer model receives a user prompt to summarize an article, it analyzes the text and generates a concise summary that only contains the article's key points.

Training with an extensive and diverse dataset enables LLMs to process many tasks in numerous areas, including healthcare and education \cite{analyticsinsight}. For instance, InteliHealth Company develops personal health LLMs to generate recommendations on personalized health plans for patients based on their medical history combined with clinical data \cite{nelsonadvisors}. In education, an interactive textbook - CourseKata using LLMs shows the benefits of training diverse datasets to meet educational needs \cite{escholarship}. Incorporating datasets from textbook materials, student responses, and assessments allows CourseKata to provide students with personalized feedback in real time via the generation of personalized practice questions by offering detailed explanations and adapting to each student's learning pace \cite{escholarship}. These examples highlight how training on large and diverse datasets enables LLMs to generalize to different domains.

\begin{figure}[h]
  \centering
  \includegraphics[width=8cm, height=2.75cm]{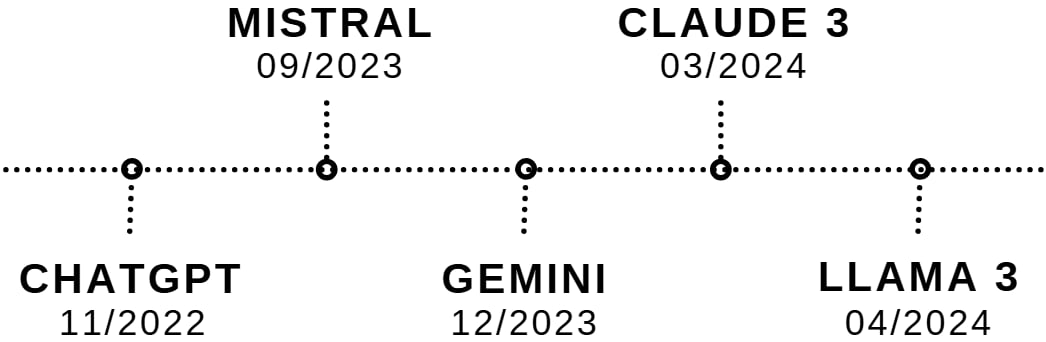}
  \label{llms_example}
  \caption{Notable released LLMs timeline}
\end{figure}

Both deep learning architectures and extensive and diverse datasets trained in LLMs lay the foundations for practical real-world applications in various industries like BERT from Google. These applications show the transformative potential of LLMs and the commitment of leading tech companies to advance their development and unlock new possibilities in various domains. For example, Google leverages BERT, with its ability to extract important information, summarize long texts, and optimize search results, to summarize texts with precision and conciseness \cite{pp}. In addition, Microsoft developed an LLM called Turing-NLG to enhance its system for identifying and extracting meaningful information from text (e.g. names, locations, and dates), allowing Microsoft to enhance language understanding, deliver reliable, context-aware information, and improve applications in NLP, search engines, and information retrieval \cite{pp}. Furthermore, IBM uses the WATSON NLU (Natural Language Understanding) model, which leverages LLMs to analyze and extract valuable insights from customer feedback, social media posts, etc., allowing them to make decisions based on this information \cite{pp}. These real-world applications illustrate how LLMs are revolutionizing industries and unlocking innovative solutions in diverse domains.

\subsection{LLMs for Code Generations} 
To begin with, we will introduce the steps of how LLMs generate code, in terms of before data feed into LLMs, during training, and after training. Before training, data preprocessing plays a crucial step in ensuring that datasets (e.g. open source repositories from GitHub and Overflow) are clean, standardized, and suitable for the chosen models, maximizing LLMs' ability to learn \cite{nvidiaexplain}. During training, LLMs build complicated internal code representations to comprehend their meanings (semantics), structures (syntax), and relationships of various code elements \cite{aiverseinfo}. Furthermore, the generating code process consists of four steps: Understanding the Prompt, Retrieving Relevant Code Patterns, Assembling Code Fragments, and Generating code \cite{aiverseinfo}. In the first step, LLMs analyze the given input by breaking it down to understand the intended functionality, the programming language, and any specified constraints. Secondly, LLMs find their input's internal representations to match the prompt's requirements by finding code patterns and structures. Subsequently, they smartly combine retrieved code fragments and modify them to fit the prompt's context in the third step. Lastly, LLMs generate final code output with many variations or suggestions for users to decide.

With the transformative process of how LLMs generate code, these models have shown rapid development in just a few years. The journey started in 2021 with GitHub Copilot, becoming one of the first widely available code generation tools \cite{infoworld1}. From then on, numerous LLMs with code generation capability have been developed. By 2022, more advanced models like Replit Ghostwriter were released, which allow users to perform tasks, such as code completion, explanation, transformation, generation, and error detection with debugging \cite{infoworld1}. In 2023, Bard was announced to support coding in more than 20 programming languages such as C++, Go, Java, JavaScript, TypeScript, and Python \cite{infoworld1}. These milestones illustrate the rapid advancements in LLMs with the capability of generating code that changes the way users approach coding tasks.

To further demonstrate the impact of advancements in LLMs on code generation, OpenAI o1, also known as ``Strawberry,'' shows a significant leap forward in coding performance. Compared to other top-performing models like 3.5 Sonnet, GPT-4o, and Llama 405b in coding challenges using HumanEval benchmark data, OpenAI o1 achieves the highest performance rate of 92.4\% which establishes itself as the best coding model according to Vellum\cite{vellum}. OpenAI o1 - one of the latest OpenAI models - presents its groundbreaking technology of ``thinking capacity'' \cite{aimlapi}. Unlike previous models that focus on the number of parameters, OpenAI o1 has the ability to ``think'' before responding by creating a long internal chain of thought, similar to how humans brainstorm to respond to problems \cite{techfundingnews}. Using a large-scale reinforcement learning algorithm, OpenAI o1 can detect and fix its mistakes, break down complicated problems into simpler components as well as attempt a new approach if the old one is not working \cite{o1}. OpenAI o1 outperforms previous models like GPT-4o by ranking in the 89th percentile on Codeforces and has the skills to solve PhD-level problems when evaluating on GPQA diamond, a difficult intelligence benchmark \cite{o1}.
  
\section{Limits and Challenges of Using LLMs for Code Generation}
The rapid development and huge potential capabilities of LLMs also raise several significant limitations and challenges. In this section, we discuss four areas: Resource Constraints, Syntactic and Semantic Errors, Bias, and Security Risks. Firstly, the training and deployment of LLMs require immense processing power and memory. For example, as one token defines one unit of text (e.g. a word or a small piece of a sentence), Llama 3.1 models were trained on over 15 trillion tokens, in which Llama 3.1 - 8B required 7 million GPU hours while Llama 3.1 - 405B required approximately 31 million GPU hours. \cite{huggingfacellama}. Secondly, LLM performance, such as accuracy and reliability, would be significantly impacted by syntactic and semantic errors, resulting in failure during execution or program generation with unexpected incorrect output \cite{prompthub}. For example, code generation can contain errors in ``if'' statements, as these can lead to incorrect branching, such as skipping conditions or executing the wrong code parts \cite{prompthub}. Thirdly, a bias testing framework called Code Bias Score (CBS) revealed that 38.92\% of GPT-4's generated code contained gender bias \cite{bias1}. Finally, security risks of LLMs may come from their training data with unsanitized open source code as an example that contains vulnerabilities such as memory safety violations or SQL injection risks\cite{security1}.

\subsection{Resource Constraints}
Chen et al. \cite{Chen} point out that training LLMs for generating code is highly resource intensive, as these models require vast computational capacity and memory, especially on Graphics Processing Units (GPUs). For instance, StarCoder - an LLM trained on over 80 PLs from GitHub - is a 15B model trained on 1T tokens. These models demonstrate significant computational demands. Recent trillion-parameter models have stretched the limits of current computational capacity for consuming extensive processing resources and memory. For example, CodeLlama, a coding model from Llama2, is available in the following model sizes: 7B, 13B, 34B, and 70B parameters. All of these, except the 70B model, have been trained on 500B tokens of code and code-related data, whereas the 70B model required 1 trillion tokens as they continue to push the limits on current state-of-the-art computing resources.

Meanwhile, Chavan et al. \cite{Nyun} highlight the critical challenges of optimizing LLMs for faster and more memory-efficient, particularly in resource-constrained environments. For example, loading a LLaMa-70B model requires 140GB of VRAM, which excludes the additional memory needed for inference. This addressed the need for model compression when deploying in a resource-constrained environment. Because of this, quantization, which reduces memory usage by lowering the numerical precision of model weights, is introduced to handle these challenges. However, various methods such as GPTQ, AWQ, OmniQuant demonstrate the trade-offs between memory efficiency and model accuracy. This study presents a performance comparison between these quantization methods in compressing the LLaMA2-7B model. For example, OmniQuant reduces memory usage by lowering the precision to 4-bit, achieving strong performance with a perplexity of 5.97. However, reducing precision further to 3-bit with OmniQuant increases the perplexity to 6.65, indicating a decline in output quality. Similarly, GPTQ with 3-bit precision reduces weight memory to 2.87 GB but leads to a higher perplexity score (7.36) compared to 4-bit (6.08) and 8-bit (5.86) configurations.

In contrast, Hassid et al. \cite{Hassid} examine the impact of computational constraints on LLM performance by comparing smaller models such as Code Llama of 7B/8B/13B with larger models such as the 70B model under the same resource limits. Interestingly, smaller models of 7B/8B/13B demonstrated better results by showing 5 to 15\% performance gains over the 70B model. In a ``small budget regime'' capped at 32 normalized FLOP units and 64 wall-time units, the study evaluated code generation benchmarks (HumanEval, MBPP) in different model sizes of 7B/8B/13B models against the 70B model. Evaluation tasks include HumanEval with 164 function completions and MBPP with 500 code generation instructions. In addition, CodeLlama 7B/13B achieved a 60\% score in only a quarter of the time required by larger models to achieve the same result on the HumanEval benchmark. This finding highlights the efficiency of smaller models and their ability to deliver competitive results with fewer resources in limited environments.

\renewcommand{\arraystretch}{2} 
\begin{table}[h]
\centering
\caption{Summary of Resource Constraints}
\begin{tabular}{|>{\centering\arraybackslash}m{0.15\columnwidth}|
                >{\centering\arraybackslash}m{0.35\columnwidth}|
                >{\centering\arraybackslash}m{0.33\columnwidth}|}
\hline
\textbf{Source} & \textbf{Key Ideas} & \textbf{Challenges} \\
\hline
\textbf{Chen et al. \cite{Chen}}  & Training inefficiency & A 15B model trained on 1T tokens. \\
\hline
\textbf{Chavan et al. \cite{Nyun}}  & Issues for faster and lighter LLMs. & Quantization method trade-offs. \\
\hline
\textbf{Hassid et al. \cite{Hassid}} & Performance comparison between smaller and larger LLMs. & 7B/13B models outperform the 70B model. \\
\hline
\end{tabular}
\end{table}

\subsection{Syntactic and Semantic Errors}
Wang et al. \cite{Wang} provide an in-depth analysis of common syntactic and semantic errors in code generated by six prominent LLMs-CodeGen-16B, InCoder-1.3B, GPT-3.5, GPT-4, SantaCoder, and StarCoder by evaluating their 557 incorrect code snippets across 164 Python tasks from the HumanEval dataset. Syntactic errors, such as missing or incorrectly structured code blocks, were two common problems with the six LLMs generating 40\% of these errors. This indicated that even widely used LLMs struggle with fundamental code structure. Simple errors, including ``if'' errors, incorrect function names, and incorrect function arguments, were also common and easy to correct. The most common semantic issues from the code generated by these six LLMs are incorrect logical flow and flawed conditional statements. In contrast to the other four models (CodeGen-16B, InCoder-1.3B, SantaCoder, and StarCoder), GPT-3.5 and GPT-4 show better performance for generating code with fewer missing steps.

Based on these findings, Dou et al. \cite{Dou} conduct a study on both syntactic and semantic errors in LLM-generated code by analyzing seven different models such as GPT-4, GPT-3.5, Claude-3, Llama-3, Phi-3, StarCoder-2, and DeepSeekCoder on three benchmarks: HumanEval+, MBPP+, and APPS+. They observed that syntactic errors, including incorrect syntax structure, indentation errors, and missing library import, are relatively rare, accounting for less than 10\% of the total errors among all models. In contrast, semantic errors, such as misunderstanding task requirements, logic errors, hallucinations, and input/output format issues, constitute the largest category of errors. DeepSeekCoder, Llama-3, Phi-3, and GPT-3.5 have proportions of semantic error that exceed 50\% on the APPS+ benchmark, showing their struggles with intricate logic and conditional structures. Additionally, as the complexity of the benchmark grows, semantic errors increase proportionally, highlighting the challenges LLMs face in accurately interpreting and executing complex tasks.

Extending the scope to translation tasks, Pan et al. \cite{Pan} categorize translation errors in LLM code into 15 types, covering both syntactic and semantic issues. Syntactic errors, often involving misalignment with target language-specific requirements, and semantic errors, which affect the logical consistency of translated code, were common. The study used real-world projects like Apache Commons CLI and Python Click to evaluate LLMs' effectiveness in code translation by categorizing translation errors and assessing the resulting syntactic and semantic issues across multiple benchmarks, such as HumanEval, MBPP, and APPS. In particular, 30.5\% of translation errors resulted from syntactic and semantic misalignments between the source and target languages, and 24.3\% of these errors involved unmet target language requirements. This study underscores the challenges LLMs face in code translation tasks, where nearly 80\% of the issues arise from such discrepancies.

Finally, Liu et al. \cite{Liu} analyze ChatGPT's correctness on code generation by examining their semantic and syntactic errors that impact code reliability and quality. The study was analyzed on 4,066 code snippets generated in Java and Python in 2,033 programming tasks, revealing that both types of errors affect the compilation and runtime errors of the generated program. The result demonstrates that Illegal Index and Type Mismatch errors are the most common semantic errors in ChatGPT-generated code. Illegal Index errors account for 46.4\% of the 97 runtime errors in Java, while Type Mismatch errors are more frequent in Python because of its dynamic typing system. Furthermore, for semantic errors, 1,930 snippets (47\%) exhibited maintainability issues, such as inconsistent variable use and improper loop handling, affecting readability and reusability. This breakdown underlines semantic and syntactic issues that lead not only to runtime errors but also to a higher demand for manual correction to achieve functional code.

\subsection{Biases}
Wang et al. \cite{bias2} explore multilingual bias in LLM code generation, including Multi-NL bias and Multi-PL bias. The paper studied multilingual bias using three popular LLMs, such as StarCoder, CodeLlama, and DeepSeek-Coder, while evaluating it on the Pass@k metric. For bias in multi-NL, the results showed that LLMs exhibit a significant performance gap when generating code from different language instructions like English and Chinese across different PLs (e.g. Python, Java, C++, etc.). Using Chinese instructions led to an average Pass@1 rate drop of 17.2\% for base models and 14.3\% for instruction-tuned models in Python, with CodeLlama-34B experiencing more severe bias as its Java code generation dropped by 37.8\%. For bias in multi-PL, the results showed various LLMs' performance in generating code in different PLs. Base models achieved the highest Pass@1 rate in Python for outperforming C++ and Java by 5.7\% and 11.3\%, respectively.

Expanding the discussion to social biases, Liu et al. \cite{bias3} investigate the severity of these biases in the generation by LLM code. Their experiments were performed on different LLMs such as Codex, InCoder, and CodeGen with different sizes to evaluate social biases in code using three metrics, such as Code Bias Score (CBS), UnFairness Score (UFS), and the standard deviation of the frequency for all valid demographics (e.g., ethnicity, religion, and gender). The results revealed that models such as Codex and InCoder generated harmful codes in which certain ethnicities or religions were associated with the derogatory term ``disgusting'' by expressing prejudice against ``Islam'' and ``Muslim''. Furthermore, Codex, with over 100 billion parameters, achieved the highest code generation quality (Pass@1: 47.03\%) but also demonstrated the most severe biases (CBS: 82.64\%), highlighting a troubling trade-off between performance and fairness. Similarly, as the sizes of CodeGen model increase from 350M to 6.1B, their performance improves from 12.76\% to 26.13\% on the Pass@1 but reveals a sharp increase in the bias of CBS escalating from 9.36\% to 62.65\%.

\subsection{Security Risks}
Islam et al. \cite{Islam2024} introduce security vulnerabilities in LLMs including three main technical issues data quality, model design, and evaluation practices. LLMs show their disadvantage in producing 10\% more vulnerable code than human developers. Data quality issues, including incorrect labeling and data leakage, as indicated by datasets such as MVD and Devign, were observed to trigger the generation of false positives or false negatives in vulnerability detection. In addition, models designed only for supervised fine-tuning, such as VulRepair, mostly generate non-functional code due to scarce syntax and functionality checks. Lastly, for evaluation, the common metrics used to evaluate these models, such as BLEU and Exact Match, are not indicative enough for the security and functionality of the generated code.

Based on this analysis, He \cite{He2023} explores recent efforts to evaluate code security by LLMs from systematic testing to user studies. Initially, the author discusses a popular security risk called ``Out-of-Bounds Write'' (CWE-787), which can allow attackers to exploit computer memory for criminal activities by writing malicious information. Recent efforts to assess the security of LLM-generated code include systematic evaluations using Common Weakness Enumeration (CWE), focusing on how Copilot handles various vulnerabilities across different prompts, weaknesses, and programming domains. Copilot's response to the scenarios of the diversity of prompt and domain shows that around 40\% of the generated code is vulnerable from a security standpoint. In addition, a security-driven user study examines code written by student programmers with LLM's assistance. The user study found that while LLM-assisted code generation introduced some vulnerabilities, the overall impact on security was small. AI-assisted students produced security-critical bugs about 10\% more often than non-assisted students.

Furthermore, Black et al. \cite{Black2024} investigate the security issues with LLM-generated code that arise from challenges in balancing security and correctness based on prompting strategies, model selection, and the degree of randomness allowed in responses. CWE-22, which is a directory traversal, and CWE-190, which is an integer overflow, have been two of the common vulnerabilities used as benchmarks to evaluate generated programs. In CWE-22 (directory traversal), the task is to generate programs that write files to specified paths. The results show that GPT-3.5 generated code that allowed filenames with ``../'', enabling unauthorized access to parent directories. In CWE-190 (integer overflow), the task required generating programs to handle numerical operations safely. The results show that Claude Opus initially used standard int types that failed to handle large numbers such as 2 * 9,999,999,999, resulting in incorrect output.

Finally, Wang et al. \cite{Wang2024} highlight the security risks in the LLM code generated in their training and during their generation process. Firstly, LLM training using unsanitized data from open source such as GitHub can lead to potential risks of inadvertently embedding security vulnerabilities in generated code. For example, the 2022 Open Source Security and Risk Analysis (OSSRA) report highlights that 81\% of the 2,049 codebases analyzed had at least one vulnerability, with 49\% contained high-risk vulnerabilities. Therefore, these models are prone to propagating vulnerabilities during the code generation process, potentially resulting in flawed outputs that are highly susceptible to exploitation and malicious attacks. To better illustrate this, Copilot generates insecure code in about 40\% of the cases, whereas ChatGPT showed that of the 21 programs, only 5 were initially secure.

\section{Fine-Tuning Techniques for Enhancing LLM Performance in Code Generation}
To handle the limitations and challenges of code generation by LLMs, fine-tuning has become an important strategy to enhance LLMs' capabilities. Fine-tuning allows users to put the pre-trained LLMs into use more for specialized applications with significantly improved performance while preserving the remaining knowledge. For example, Google revealed that the fine-tuning of sentiment analysis boosted the accuracy of LLMs by 10\% \cite{turing}. Because of that, this section explores three fine-tuning techniques: prompt engineering, which optimizes LLM outputs by crafting effective input instructions; feedback refinement, which reduces errors by incorporating corrections; and domain-specific dataset tuning, which improves LLM performance in specialized areas. Together, these techniques mitigate specific weaknesses within LLMs and open the way to more effective and robust applications involving code generation.

\begin{figure}[h]
  \centering
  \includegraphics[width=6.5cm, height=5cm]{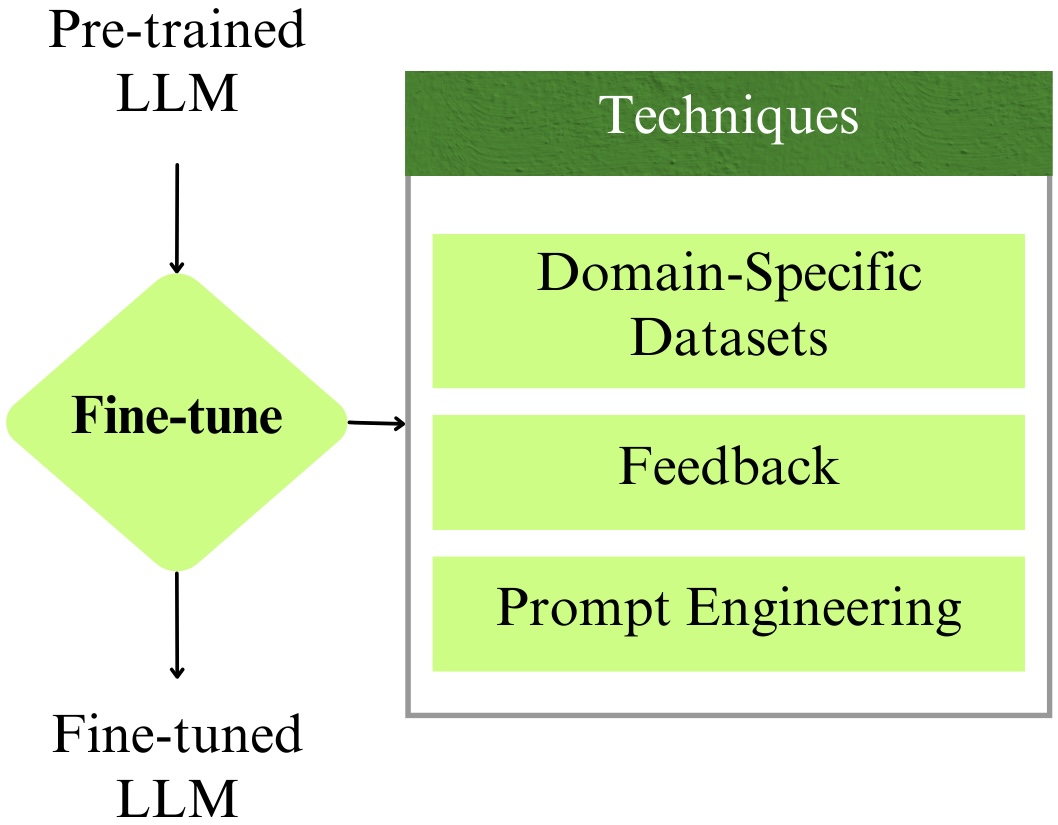}
  \caption{LLMs-Based Fine-tuning process}
  \label{finetune_process}
\end{figure}

\subsection{Fine-Tuning on Domain-Specific Datasets}
Ma et al. \cite{llamoco} propose LLaMoCo, the first instruction-tuning framework designed to adapt LLMs for the optimization of code generation. This framework establishes a comprehensive instruction set that contains well-described problem prompts and effective optimization codes. Then, it proposes a new two-phase learning strategy that includes a contrast learning-based warm-up procedure before the instruction-tuning phase to boost the convergence behavior during model fine-tuning. The results of the experiment showed that LLaMoCo significantly improved the performance of LLMs as a fine-tuned CodeGen (350M) model of LLaMoCo demonstrated superior optimization performance compared to GPT-4 Turbo on both synthetic and realistic problem sets, showing less error for 4.168\% and 79.483\% and better performance for 87.227\% and 59.174\%, respectively. In addition, LlaMoCo boosted CodeLlama 7B model performance from 29.717\% to 81.843\%.

Furthermore, Weyssow et al. \cite{4peft} review four popular Parameter-Efficient Fine-Tuning-LoRA, IA3, Prompt Tuning, and Prefix Tuning. These techniques ensure fine-tuning of LLMs by updating only a subset of model parameters rather than all the parameters. Thus, LLM focuses on task-specific data while maintaining good resource usage. Then, this paper compares these four techniques with ICL and traditional full fine-tuning on code generation tasks using Python datasets like CoNaLa or CodeAlpaca. The results indicated that the fine-tuned LLMs consistently perform significantly better with PEFT compared to ICL as LoRA fine-tuning LLMs improved 25.4\% and 22.8\% (150\% and 29.8\%) in EM@10 and CodeBLEU respectively in CoNaLa (CodeAlpacaPy) dataset. QLoRA reduces memory usage by allowing fine-tuning of LLMs with up to 34B parameters. This investigation emphasizes the potential of PEFT techniques in efficiently fine-tuning LLMs to task-specific data to generate code.

Complementing these approaches, Tsai et al. \cite{datapruning} introduce a novel approach to fine-tuning LLMs for code generation by integrating data pruning methods. The paper explores the use of clustering algorithms (KMeans, Agglomerative Clustering, HDBSCAN) and pruning metrics (Diversity Metric, Density Metric) to reduce the size of training data selectively while maintaining the accuracy and functionality of the generated code as there are significant redundancies in training data. HumanEval(+) and MBPP(+) datasets are used to evaluate pruning methods and highlight performance improvements. Surprisingly, the results show that pruning in a small portion of the training data can lead to performance improvements of up to 2.7\% in HumanEval and 3.5\% in MBPP. Remarkably, using data pruning on only 1\% of the data can result in a 4.1\% improvement compared to the base model, achieving performance nearly equivalent to training with the entire dataset.

\subsection{Feedback}
Mu et al. \cite{mu} present a novel framework - ClarifyGPT - that can identify and clarify ambiguous user requirements to improve LLM-based code generation. ClarifyGPT can perform a code consistency check to detect ambiguity and generate targeted clarifying questions to refine unclear input. Consequently, it generates the solution code from the received response. Therefore, this framework plays an important role in improving the interpretability of the code generated by LLMs. Furthermore, it helps users better understand generated code from interaction and provides more clarification of their intentions. Using two publicly available benchmarks: MBPP-sanitized and MBPP-ET for evaluation, ClarifyGPT improved the average performance of GPT-4 and ChatGPT from 68.02\% to 75.75\% and from 58.55\% to 67.22\%, respectively.

Furthermore, Gehring et al. \cite{gehring} discuss Reinforcement Learning from Execution Feedback (RLEF), a method to improve LLMs in code synthesis by using feedback from code execution to iteratively refine outputs. The process includes three steps: generating code, receiving feedback from test cases, and updating the model through reinforcement learning using Proximal Policy Optimization (PPO). In experiments on competitive programming tasks such as those in CodeContests, models that are trained with RLEF achieved a solve rate of 37.5\% in the test set for the standalone Llama 3.1 70B model. These significantly outperform the previous state-of-the-art AlphaCodium at 29\%. The method also reduced samples by an order of magnitude compared to the RLEF approach. This approach generalizes well to other benchmarks like HumanEval+ and MBPP+, where feedback is used for the grounding of the output of LLMs, especially on multi-turn code generation tasks.

Finally, Wong et al. \cite{cRLHF} introduce a new method that combines crowd-sourced computation and reinforcement learning from human feedback (cRLHF), to improve code generation in LLMs. This aims to maximize code quality using multiple user feedback. As the traditional method - RLHF contains biases and misses important insights that limit LLMs' potential, cRLHF collects feedback data from different sources and uses Bayesian inference to align and combine the feedback data into one belief that gives more objective assessments without complicated reward modeling. The framework fine-tunes LLMs by using aggregated feedback to improve code correctness and quality. The results show significant improvements in LLMs of different sizes when the cRLHF method is applied. In the HumanEval benchmark evaluation, the success rate for CodeGen-2.7B improved from 39.8\% to 45.4\% and from 17.3\% to 20.0\% for the smaller model CodeGen-350M.

\subsection{Prompt Engineering}
Sun et al. \cite{sun2024} apply the ``Chain-of-Thought'' prompting technique to generate ``solution plans'' for complex programming challenges to develop a framework called CodePLAN. This framework is designed to infuse the reasoning capabilities of LLMs in smaller models to enhance their code generation performance. CodePLAN uses multitask learning to train smaller models on both code generation and solution plan generation simultaneously. It uses backward reasoning and plan sampling strategies to improve solution plan quality. The higher quality of the solution plan may lead to more accurate code generation outputs. The framework considers LLMs as ``teachers'' to provide solution plans that distill into smaller models considered ``students''. This allows them to develop solution plans independently at inference time. Experiments demonstrated that this approach significantly improves the code generation abilities of smaller models by more than 130\% in performance using the pass@1 metric on the APPS benchmark.

Expanding on prompting techniques, Li et al. \cite{li2024} develop a novel approach named AceCoder to improve LLM's performance in code generation. It is designed to perform two major challenges of code generation: requirement understanding and code implementation. This method performs code generation in three steps: example retrieval, prompt construction, and code generation. First, the retriever selects similar programs based on language input, whereas the selector selects non-redundant programs based on prioritizing non-overlapping information. Second, the technique identifies a combination of chosen examples, their preliminary artifacts in the form of test cases, and input requirements to construct a prompt. Finally, the LLM uses the constructed prompt to generate test cases that yield the final source code. AceCoder was evaluated on three LLMs, such as CodeGeeX, CodeGen, and InCoder, using three public benchmarks using Pass@k. It follows that AceCoder has surpassed the state-of-the-art prompting techniques in improving Pass@1 by up to 56.4\% in MBPP, 70.7\% in MBJP, and 88.4\% in MBJSP and has proven to be effective in different LLM sizes and languages.

Lastly, Tony et al. \cite{5prompt} explore the impact of different prompting techniques on the security of the code generated by LLMs from NL instructions. These techniques were implemented in the GPT-3, GPT-3.5, and GPT-4 models. The authors investigated some of these techniques using a dataset of 150 NL prompts related to security-relevant coding tasks. 15 different explored prompting techniques are classified into 5 categories depending on their common characteristics, such as root techniques, refinement-based techniques, decomposition-based techniques, reasoning-based techniques, and priming techniques. For instance, refinement-based techniques focus on improving model outputs through iterative refinement, feedback loops, or self-assessment, including methods such as Recursive Criticism and Improvement (RCI), Self-refine, and Progressive Hint prompting. The results indicated that RCI performed best for both GPT-3.5 and GPT-4, while zero-shot prompting performed best out of these techniques for GPT-3. The persona / memetic proxy yielded the poorest performance, generating the most security weaknesses across all models.

\section{Evaluation Metrics and Benchmarks for Assessing LLM-Generated Code}
As numerous LLMs with code generation capability have been developing - a crucial tool for programmers of all skill levels, evaluating these models is essential to ensure their dependability and efficiency in meeting users' needs. While significant efforts have been dedicated to the performance evaluation of LLMs, most of these research questions remain unanswered, such as: ``Are the evaluations and comparisons fair and are the differences significant?'' or ``Do findings from performance evaluation truly reflect the usability of LLMs as practical programming tools?'' \cite{paul2024}. This section will discuss two key aspects of evaluation: benchmarks and performance metrics.

\begin{figure}[h]
  \centering
  \includegraphics[width=7.5cm, height=5cm]{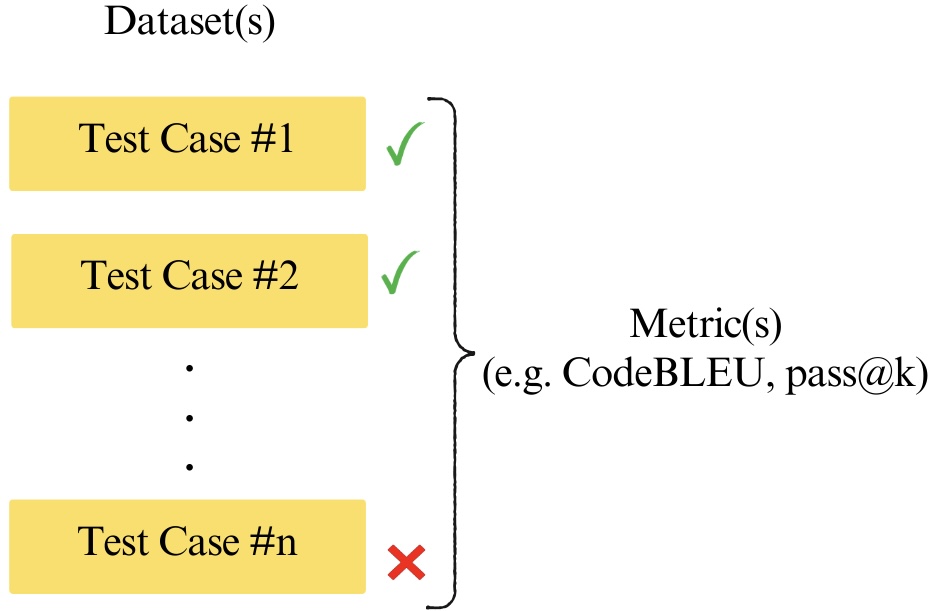}
  \caption{LLMs System Benchmark \cite{benchmark}}
  \label{benchmark_structure}
\end{figure}

Before we discuss these aspects in depth, it is essential to clarify the terms ``benchmarks'' and ``metrics''. LLM evaluation metrics are criteria used to quantify the performance of LLM systems in aspects such as correctness of the answers, semantic similarity, and hallucination \cite{metric}. On the other hand, benchmarks are constructed from evaluation datasets and metrics where test cases create an evaluation dataset \cite{benchmark}. Figure \ref{benchmark_structure} illustrates the LLM benchmark structure that includes the integration of metrics within this evaluation framework.

\subsection{Metrics}
One common metric for evaluating code generation by LLMs is CodeBLEU \cite{Ren2020}. Compared to the generally adopted BLEU metric for NL evaluation, which lacks key syntactical and semantic characteristics of codes, CodeBLEU was designed to incorporate both traditional n-gram matching and syntactic and semantic matching. Specifically, the n-gram match weights different n-grams differently, the syntactic match plugs in AST information by aligning subtrees, while the semantic match measures similarities of code based on the analysis of its data-flow structure. CodeBLEU combines these elements (including weighted n-gram match, syntactic AST match, and semantic data-flow match) into a comprehensive evaluation metric. The experiments are tested in three coding tasks text-to-code (Java), code translation (from Java to C\#), and code refinement (Java). The results demonstrate that CodeBLEU has a better correlation with human evaluation scores compared to traditional metrics like BLEU and perfect accuracy in all three tasks.

Another popularly used metric for LLMs generated code is pass@k \cite{humaneval}. Pass@k, with the goal of addressing the shortcomings of traditional text similarity metrics, is designed to assess the functional correctness of generated code samples. It presents the probability that at least one of the top k samples passes unit tests. This metric contains variations, such as pass@1, pass@10, pass@100, etc. Pass@1 provides the likelihood of correctness at the first attempt, while pass@10 and pass@100 assess the model's performance on much larger sample sets to provide a comprehensive view of its ability to generate valid solutions. The following formula is used to calculate pass@k to handle all problems with E defining the expected value of the problems, n for the total number of samples, and k for the number of top samples to consider:
\[
\text{pass@k} := \mathbb{E}_{\text{problems}} \left[ 1 - \frac{C(n-c, k)}{C(n, k)} \right]
\]

Yeo et al. \cite{Yeo2024} propose a new metric called the pass-ratio@n that measures precision based on the accuracy granularity through the pass rate of the test cases. As LLMs can generate different solutions across inferences, considering that n inferences are made, the average pass-ratio across the n solutions is used to mitigate bias. For each solution $i$ ($0 < i \leq n$), the pass-ratio is calculated by the following formula:
\[
\textit{pass-ratio}_i = \left( \frac{\text{\# of test cases passed at code } i}{\text{\# of test cases}} \right)^2
\]\\
And the pass-ratio@n demonstrates the mean pass-ratio of n generated code.
\[
\textit{pass-ratio}@n = \frac{\sum_{i=1}^{n} \textit{pass-ratio}_i}{n}
\]
The metric pass-ratio@n was tested on three coding problems and five generated coding solutions for each problem from LLM inference to compare to pass@k. The results show that the pass-ratio@n can provide more granular insight than the pass@k metrics. In a coding problem where none of the solutions passed all test cases, pass-ratio@5 got a partial score of 61\%, while pass@k recorded 0\%.

In addition, Zhuo \cite{zhuo} introduces another new metric called ICE-Score by instructing LLMs for code assessments. The ICE-Score metric includes two key components: (1) the task definition, evaluation criteria, and structured evaluation steps, and (2) the provided problem along with the generated code snippet for assessment. Unlike traditional metrics such as BLEU, reference-based methods relying on human-written test suites, ICE-Score uses LLMs' capabilities to assess generated code on two aspects: usefulness and functional correctness. By entering the task problems and their generated code, ICE-Score outputs the corresponding assessments. The results outputted for both aspects include ``Nearly Useless'' or ``Totally Useless'', and ``Functional Incorrect'' or ``Functional Correct''.

\subsection{Benchmarks}
Table \ref{benchmark_summary} represents the information of several notable benchmarks, which will be discussed in this benchmark section from Symflower \cite{symflower}.

\renewcommand{\arraystretch}{2.5} 
\setlength{\tabcolsep}{5pt} 
\begin{table}[h]
\centering
\caption{Benchmark Features Summary}
\label{benchmark_summary}
\begin{tabular}{|>{\centering\arraybackslash}m{0.3\columnwidth}|
                >{\centering\arraybackslash}m{0.17\columnwidth}|
                >{\centering\arraybackslash}m{0.13\columnwidth}|
                >{\centering\arraybackslash}m{0.25\columnwidth}|}
\hline
\textbf{Benchmarks} & \textbf{Number of Tasks} & \textbf{PLs} & \textbf{Release Date} \\
\hline
HumanEval & 164 & Python & July 2021 \\
\hline
CLASSEVAL & 100 & Python & August 2023 \\
\hline
SWE-bench & 2,294 & Python & October 2023 \\
\hline
BigCodeBench & 1,140 & Python & June 2024 \\
\hline
\end{tabular}
\end{table}

The HumanEval \cite{humaneval} dataset is a benchmark designed to evaluate LLMs in code generation. It includes 164 programming challenges as each problem contains ``function signatures, docstrings, body, and unit tests'' to evaluate functional correctness. On average, there are 7.7 tests per problem. Traditional metrics like BLEU can measure the similarity of texts. Unfortunately, these measures are not suitable for evaluating code generation because functional correctness is much more important. This was addressed by introducing the pass@k metric, which came with the HumanEval dataset and helped to assess this functional correctness. This metric measures the probability that at least one of the top k-generated code samples passes the unit tests to provide a more practical evaluation of the generated code. HumanEval and pass@k have become critical factors in testing LLM coding capabilities to provide more meaningful and valuable test results.

In contrast, ClassEval \cite{classeval} is a benchmark designed to evaluate LLMs on the more challenging coding tasks, such as class-level code generation, unlike existing benchmarks like HumanEval that focus only on simple scenarios, such as function-level or statement-level. The benchmark consists of 100 manually constructed Python coding tasks, including 100 classes and 412 methods. The study experimented with the evaluation of 11 state-of-the-art LLMs in class-level generated code using three different code generation strategies, including holistic generation, incremental generation, and compositional generation. First, existing LLMs perform significantly worse in class-level code generation compared to standalone method-level benchmarks like HumanEval. Secondly, GPT-4 and GPT-3.5 consistently outperform other models; and models such as WizardCoder, Instruct-StarCoder, and Instruct-CodeGen have similar performance. Lastly, the best strategy for GPT-4 and GPT-3.5 is to generate code for the entire class, while using a method-by-method strategy will be a better choice for other models.

Furthermore, SWE-bench \cite{swe} is a benchmark designed to evaluate LLMs for the study of capabilities in real-world software engineering settings. As an evaluation framework, this benchmark comprises 2,294 tasks from Github issues and their related pull requests across 12 well-known Python repositories. Solving issues in the SWE-bench often requires a comprehensive understanding and coordination with changes across various functions, classes, and even multiple files at the same time, requiring models to interact with execution environments, handle extensive contexts, and carry out complex reasoning beyond typical code generation tasks. In addition, the evaluation experiment revealed that both advanced proprietary models and the fine-tuned model of the paper, SWE-Llama, can handle only the most simple issues. The best-performing model, Claude 2, can solve only 1.96\% of the tasks with the BM25 retriever. Therefore, SWE-bench reflects the real-world coding environments to create solutions immediately applicable in open source software development.

Finally, BigCodeBench \cite{bigcodebench} is a new benchmark designed to evaluate LLMs on tackling practical and complex programming tasks to ensure no data contamination. Due to many library and function calls in real-world software development, concerns are raised for HumanEval, which is a simpler benchmark and is affected by contamination and overfitting problems. BigCodeBench includes 1,140 function-level tasks that require LLMs to use various libraries and compose multiple function calls. On average, there are 5.6 test cases per task with branch coverage of 99\%. The benchmark tests performance using the Pass@1 metric with greedy decoding to measure the percentage of tasks correctly solved by the first generated code against curated test cases. For experiments, BigCodeBench ensures the quality of the task through collaboration between GPT-4 and 20 human experts, refining the tasks with test cases in a sandbox environment. The tasks are further evaluated by other LLMs and cross-checked by 7 experts, with the resulting average human performance at 97\%.

\section{LLMs' Applications in Code Generation and Development}
LLMs have transformed code generation in software development as these models provide user assistance in a variety of coding tasks, such as code completion, code translation, etc. In addition to that, each model has unique strengths and developers must understand and manipulate them effectively in their workflows. Selecting a suitable model for the task can give users the advantages of enhancing productivity, streamlining processes, reducing errors, and maximizing its potential. For example, OpenAI Codex benefits users whose workflow depends on GitHub due to the integration of Codex with GitHub Copilot \cite{autogpt}. Furthermore, for developers working on AWS, CodeWhisperer provides domain-specific insights and customized recommendations that position it as one of the top LLMs for cloud computing-focused development \cite{autogpt}. Moreover, tools have been designed to augment LLMs' capabilities. For example, CodeAgent - an LLM-based agent framework - was developed to assist and allow LLMs to handle complicated programming tasks \cite{codeagent}. The following sections demonstrate several aspects of the broader code generation topic and are categorized into three groups of tasks: (A) code generation and code completion (foundational tasks), (B) code generation and code search (advanced tasks), and (C) debugging and code translation (auxiliary tasks).

\subsection{Code Completion and Code Generation}
Xu \cite{copilot} introduces an advanced code autocompletion tool, GitHub Copilot, from the collaboration between GitHub's vast software development resource and OpenAI's groundbreaking AI development. GitHub Copilot uses deep learning models (recurrent neural networks (RNNs)) and transformers (Transformer) which help the model to learn the code's syntactic and semantic structure and developers' coding habits. In addition, it is capable of generating context-sensitive code suggestions from the training data of multiple open source code libraries and developers' code contributions. GitHub Copilot analyzes the code context from developers, including the currently written code (functions, classes, and methods) to understand developers' demands. Subsequently, based on given inputs, it generates code suggestions in real time and can continuously modify them depending on what developers need. This allows developers to learn from the coding community's knowledge and experience as well as enhances the reusability, quality, and efficiency of the code. Furthermore, another advantage to mention is that GitHub Copilot can save developers a lot of time and effort by reducing manual code written from repetitive or boilerplate tasks.

Moreover, Meta AI \cite{codellama} releases a state-of-the-art LLM - Code Llama for code generation built from LLaMA 2 architecture, which excels in code completion tasks. Three available models are Code Llama - the base model, Code Llama - Python for optimizing Python Programming, and Code Llama - Instruct fine-tuned for better understanding and responding to NL instructions. There are four Code Llama sizes for 7B, 13B, 34B, and 70B parameters, respectively, as each model is trained on 500B code tokens, and related data with the 70B model are specifically trained on 1T tokens. The base and instruct models with 7B and 13B are trained using fill-in-the-middle (FIM) capability. This enables models to insert code within existing code to support tasks like code completion. Additionally, smaller 7B and 13B models are quicker and more suitable for tasks requiring low latency like real-time code completion. Code Llama was tested for its ability to complete code with given docstrings using the HumanEval benchmark. The results show that Code Llama surpassed open source, code-focused LLMs and outperformed Llama 2 as model 34B achieved 53.7\% on HumanEval, matching ChatGPT in performance.

Lastly, Wang et al. \cite{toolgen} develop ToolGen to improve repository-level code generation in the LLM generation process by integrating autocompletion tools. It includes two phases: Offline Trigger Insertion and Model Fine-tuning, and Online Tool-integrated Code Generation. ToolGen tackles problems in code generation for dependency errors, such as undefined-variable and no-member errors by manipulating autocompletion tools to fill repository-level dependencies. In the experiments, ToolGen was applied to three different LLMs-CodeGPT, CodeT5, and CodeLlama, and tested on two datasets CodeSearchNet and CoderEval to evaluate similarity-based and dependency-based effectiveness, and execution-based effectiveness, respectively. The results demonstrated models' improvement by enhancing 31.4\% to 39.1\% in Dependency Coverage, and 44.9\% to 57.7\% in Static Validity Rate for the three LLMs, while maintaining competitive performances on metrics BLEU-4, CodeBLEU, Edit Similarity, and Exact Match. In addition, ToolGen improved CodeT5 and Code Llama by 40\% and 25\%, respectively, and maintained the same pass rate for CodeGPT.

\subsection{Code Search and Advanced Code Generation}
Code search is an essential task in software development practices, allowing developers to efficiently create solutions to problems. An LLM-assisted tool that can enhance this task is RepoRift. Jain et al. \cite{reporift} introduce this advanced code search application RepoRift designed to improve code snippet retrieval using LLMs with Retrieval Augmented Generation (RAG). It enhances user queries by injecting more context from GitHub repositories to address issues like ambiguity and Vocabulary Mismatch Problems. RepoRift utilizes a multi-stream ensemble architecture that refines the search results by doing multiple comparisons and generating the most relevant snippets. For evaluation in the CodeSearchNet dataset, RepoLift significantly outperformed other code search methods by successfully achieving a success rate of 78.2\% and 34.6\% at Success@10 and Success@1, respectively. Furthermore, it delivers high accuracy using minimal preprocessing of the evaluation set and efficiently manages queries in different forms.

Extending code search capabilities, Feng et al. \cite{codebert} present CodeBERT which is a bimodal pre-trained model designed to understand and generate NL and PL code, such as Python, Java, etc. With the Transformer-based neural architecture and training on a hybrid objective function combined with the pretraining task of replaced token detection, this allows codeBERT to leverage both bimodal (NL-PL pairs) and (only NL or PL) unimodal data. Based on these, CodeBERT shows its strong potential in code search. For evaluation, CodeBERT is tested using a dataset for NL-PL probing including NL code search in a zero-shot scenario and compared with an NL-based pre-trained model called RoBERTa. In an experiment on the CodeSearchNet corpus, CodeBERT performed better and more consistently than RoBERTa. Moreover, on the documentation generation task in six PLs, CodeBERT outperformed RoBERTa by achieving a 1.3 BLEU score gain and state-of-the-art performance.

Switching to advanced code generation, Li et al. \cite{alphacode} develop a model named AlphaCode to handle competitive programming problems that require advanced problem solving skills. It is initially pre-trained on selected GitHub code and fine-tuned on a curated dataset of competitive programming problems like CodeContests. The approach is to automatically generate millions of code examples, filter them according to their execution results, and cluster them, after which a small number of high-quality submissions are manually selected. For evaluation, using simulation on the Codeforces platform, AlphaCode's performance reached the top 54.3\% among more than 5,000 human competitors. To improve this model, DeepMind developed a new dataset for training and evaluation called CodeContests. It combines data from multiple sources, where training data predate the evaluation problems, adds additional tests for accuracy, and has the evaluation of submissions done in a competitive programming-like setting. This results in 34.2\% of the long-held competitive problems from CodeContests being solved by the best model. Finally, for the model's good and reliable performance, the paper found the following critical components: a high-quality competitive programming dataset, efficient transformer models, and large-scale sampling.

\subsection{Code Translation and Code Debugging}
Hou and Ji \cite{gpt4} discuss the fact in a study that GPT-4 is the top-performing LLM in generating programming code that outperforms other LLMs, such as Gemini Ultra and Claude 2. It has gained success with various forms of programming tasks, including assisting in writing code, learning from coding error messages, and code translation. In a LeetCode and GeeksforGeeks coding contest between human programmers and LLMs, the GPT-4 success rate reached over 90\% for tasks that only more than 20\% human participants could solve. These showed that GPT-4 has the ability to be a reliable coding assistant. Furthermore, using prompt strategies, GPT-4 demonstrated its ability to learn from past errors by salvaging over 60\% of easy and medium tasks after failing in the first attempt. Finally, for the task of translating the correct Python3 code to multiple different languages, it translated most of the tasks accurately. Surprisingly, in several medium tasks, it even tackled the programming task correctly despite giving the incorrect original Python3 code, proving it a reliable tool for code translation.

Furthermore, Prenner et al. \cite{codex} investigate Codex's ability to detect and fix bugs, which are essential tasks for automated program repair. Codex, built on GPT-3 architecture, has shown great potential in generating code from NL descriptions. In this paper, Codex's ability to fix software bugs was evaluated on the QuixBugs benchmark, which contains 40 bugs in both Java and Python, then comparing its performance with three APR approaches, such as CoCoNut, DeepDebug, and CURE. The results show that Codex performed the tasks surprisingly competitively, especially in Python, with 50\% more bugs fixed compared to Java despite not being trained on Automatic Program Repair (APR). Codex outperformed both CoCoNuT and DeepDebug in Python and even outperformed CoCoNut in Java. Additionally, Codex performance was also tested using different prompt strategies for bug localization and repair, revealing that prompts can significantly impact Codex's capability of fixing bugs effectively.

Finally, a notable application that can greatly assist LLMs in code translation is Flourine. Flourine \cite{flourine}, which is an end-to-end translation tool, ensures translation validation based on cross-language differential fuzzing without requiring any test case to check input-output similarity for the original and translated code. Flourine implements the feedback strategy that provides input to the LLMs, allowing it to correct the identified counterexample. Experiments were carried out on 8160 code translations of 408 code samples, four feedback strategies, and five LLMs, including GPT4, Claude 3, Claude 2.1, Gemini Pro, and Mixtral. Benchmarks are collected as real-world projects from GitHub, using C and GO as the source code. The results revealed that the most successful LLM can achieve up to 47\% of the benchmarks.

\section{Conclusion}
This survey provides an overview of the recent landscape of LLMs for automatic code generation. To begin with, we point out the limits and challenges LLM has faced, such as resource constraints; syntactic and semantic errors; biases; and security risks, highlighting factors that need to be mitigated. Subsequently, we discuss various fine-tuning techniques, including prompt engineering, reinforcement learning, and domain-specific dataset tuning which are essential approaches to handle the issues and enhance model performance and adaptability. We then examine the importance of evaluation metrics and benchmarks, as they are critical for assessing the effectiveness and reliability of the models, techniques, and their generated code to guide future development. Finally, we explore the significant potential of LLMs in many different coding tasks, including code generation, completion, search, debugging, and translation, which significantly boost productivity and efficiency for users in writing code.

\end{document}